\documentclass[aps,prl,twocolumn,groupedaddress,showpacs]{revtex4}

\usepackage{graphicx}

\begin{document}

\preprint{Taipex-AIEcon 12-31-04}

\title{Statistical properties of an experimental political futures market}


\author{Sun-Chong Wang}
\email{scwang@phys.sinica.edu.tw}
\altaffiliation{Author to whom correspondence should be addressed}
\author{Sai-Ping Li}
\email{spli@phys.sinica.edu.tw}
\affiliation{Institute of Physics, Academia Sinica, Taipei 115 Taiwan}
\author{Chung-Ching Tai}
\affiliation{Department of Economics, National Chengchi University, Taipei 116, Taiwan}
\author{Shu-Heng Chen}
\email{chchen@nccu.edu.tw}
\homepage[Website: ]{http://www.aiecon.org}
\altaffiliation{Artificial Intelligence Economic Research Center, National Chengchi University, Taipei 116, Taiwan}
\affiliation{Department of Economics, National Chengchi University, Taipei 116, Taiwan}


\date{\today}

\begin{abstract}
A 24-hour exchange market was created on the Web to trade political futures 
contracts using fictitious money. In this online market,
a political futures contract
is a futures contract which matures on the election day with a liquidation 
price determined by 
the percentage of votes a candidate receives on the election day. 
Continuous double auctions were implemented as the system for 
order storage and price discovery.
We drew market participants in the form of tournaments in which top traders won
cash awards. Such a market was run, with about 400 registered traders, 
during the 
U.S. presidential election in November 2004 and 
Taiwan parliamentary election in December 2004. 
The experiments recorded transaction price, highest bid, lowest ask, and
trading volume of each contract as a function of time. 
Despite the relatively small scale of the exchange, 
in terms of the number of participants and 
duration of the tournament, we report evidence for asymptotic
power-law behaviors of the distributions of 
price returns, trading volumes, inter-transaction time intervals, 
and accumulated
wealth that were 
found universal in real financial markets. 
\end{abstract}


\maketitle

\section{Introduction}
Markets are a complex system that usually consists of the following different 
types of 
participants: (i) producers who provide goods, 
(ii) speculators or hedgers who, with beliefs in the trends of price movements,
buy low and sell high for a profit or insurance, 
and (iii) arbitrageurs who buy products
at a low price in one market and sell them at a high price in other
markets for a riskless profit. A market is liquid 
if sufficient numbers of the 
different types of players exist in symbiosis. To study the system,
successive movements in observables such as price are modeled as a 
stochastic process due to market's responses to the random arrivals of 
information. Fluctuations are predicted to be Gaussian\cite{bachelier00} 
or L\'{e}vy\cite{levy37} distributed
by the central limit theorem. Distributions of large price changes, those which
exceed, say, five standard
deviations, however show characteristic power-law behaviors\cite{mandelbrot63,fama65,dacorogna93,loretan94,lux96,pagan96,bouchaud98,arneodo98,gopikrishnan99,plerou99}.  
Models to explain the power-law range from systems at the state of 
self-organized criticality
exhibiting scale free properties in the order parameters\cite{bak92}, to
evolutionary systems\cite{ponzi00} whose constituents interact through a social network\cite{lux99,cont00,gabaix03}.

In an attempt to experimentally study market behaviors, we created 
an online marketplace
that hosts the three types of market players mentioned above\cite{wang04}. 
In this market, a player, after free registration for an account on 
our exchange server\footnote{http://socioecono.phys.sinica.edu.tw}, 
was allocated a fixed (and common)
amount of fictitious money to start with.
We defined the so-called political futures contracts\cite{forsythe92,berg03} 
and held trading tournaments that gave cash awards to
those who fared well in their account wealth at the end of the
tournament\cite{wang04}. 
A political futures contract, say {\it Bush-Cheney}, 
is a futures contract whose 
liquidation price is set by the percentage of
(electoral College) votes the Bush-Cheney ticket receives on November 2, 2004,
when the contract matures.
A player who believes George W. Bush would win the election would buy in shares
of {\it Bush-Cheney} when the market price of a {\it Bush-Cheney} is low
(e.g. below
50). In addition to {\it Bush-Cheney} and {\it Kerry-Edwards} futures
contract, we also issue
{\it Others} to account for votes for independent candidates.
The sum of the price of each 
share of {\it Bush-Cheney}, {\it Kerry-Edwards}, and 
{\it Others} is 100 if the market is rational, 
deviations from 100 of the sum
at any time providing opportunities 
for arbitrageurs.
Players submit bid or ask limit (or market) orders online
which are matched
at real time on our server by the mechanism of continuous double auctions\cite{smith03} which 
is widely used in real world financial exchanges.
The design of tournament is aimed at 
recruiting serious participants who are believed to
make prudent decisions when   
they have a stake in the engagement.

Two tournaments were launched for anyone who had access to the Web. 
The first, between October 4 and November 3 of 2004, was on the 2004
U.S. presidential election while the second, between November 11 and December 12
of 2004, on the 2004 Taiwan parliamentary election\footnote{http://socioecono.phys.sinica.edu.tw/exchange/announce}. The exchange server, open
24 hours a day 7 days a week, recorded data including the
transaction price, volume traded, highest bid,
and lowest ask with time of each contract. The result shows a scaling property
in the probability densities 
of price returns over a range of time lags $\tau$ across 2 orders of magnitudes
($55\ {\rm min}<\tau<8103\ {\rm min}$). 
The
central region of the densities can be described by a Cauchy distribution
(a stable L\'{e}vy distribution which decays slowly
as a power law of an exponent 2)
while the tails by a 
power law, the exponent of which depends on whether transaction prices
or means of the bid-ask spread are used in obtaining the densities. 
The distribution of changes in trading volume was found to follow a
Gaussian distribution while that of large
trading volume can be fitted by a power law.
The distribution of players' wealth, which started from a delta function,
was found power law distributed when the tournament
ended. The distribution of inter-transaction time intervals was also 
found to follow a power law.
Despite the fact that the money is fictitious and the scale of
the exchange is small in terms of the number
of players and time span,
the results reproduced many properties characteristic of real financial markets.
If we consider a tournament as an experiment, 
by observing changes in the statistical properties of the market observables 
with changes in rules of the exchange, we expect the platform to shed
light on the principles that govern socioeconomic behaviors. 

\section{Experimental Design}
\begin{figure}
\includegraphics{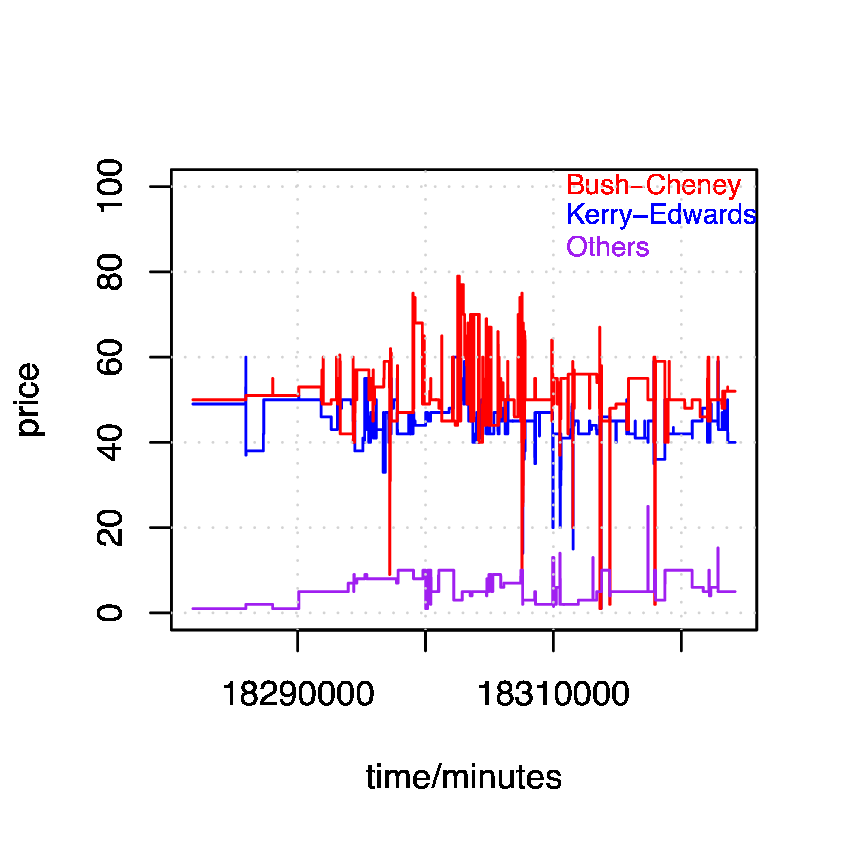}%
\caption{Price time-series for the futures in the 2004 U.S. presidential
election. Others represents votes received by 
all candidates others than Bush-Cheney and
Kerry-Edwards.}
\end{figure}

\begin{figure}
\includegraphics{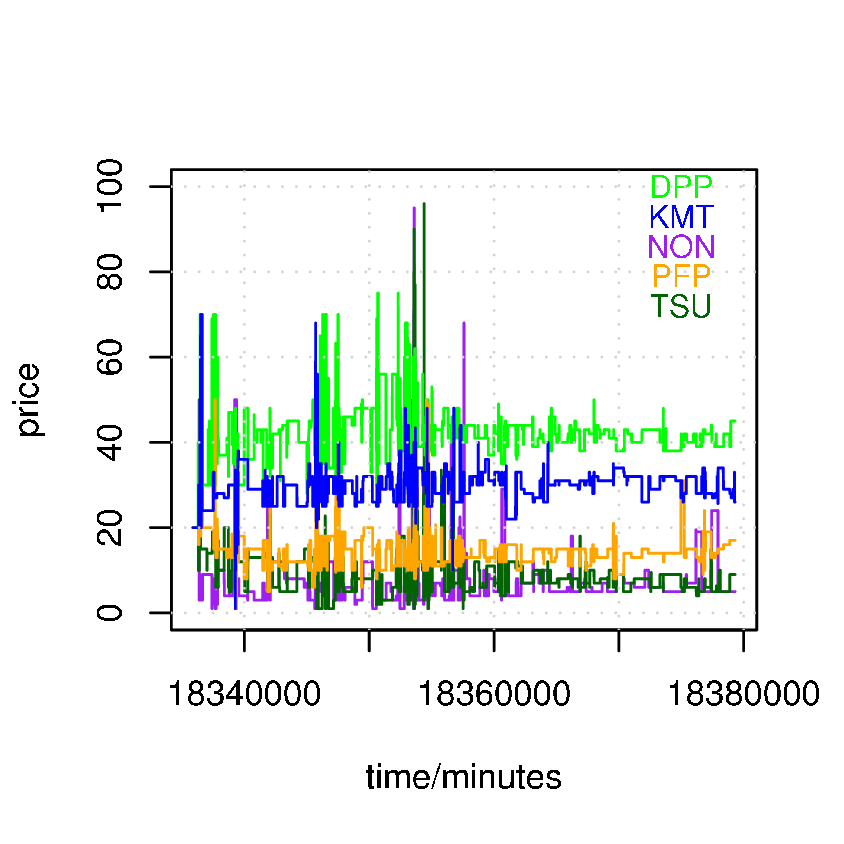}%
\caption{Price time-series for the futures in the 2004 Taiwan parliamentary
election. Labels, except NON, are the acronyms of the major political parties 
participating in the election. NON stands for all candidates other than the
four major parties.}
\end{figure}

Due to the nature of the futures, tournaments were scheduled
to start 
one month before the election day and ended on the election day
when the futures matured.
Recruiting as many players as possible presented a great challenge
to researchers who lacked marketing channels. What was done was to 
post news about the tournament to college campus bulletin boards throughout 
Taiwan\cite{wang04}.  
Numbers of registrations increased with time and reached
364 and 498 respectively for the U.S. presidential and the Taiwan parliamentary
election near the end of the tournaments. 
There was a change in the rules between the two experiments.\footnote{http://socioecono.phys.sinica.edu.tw/exchange/faq}
In the U.S. case, submitted orders waited in the orderbook for
matching orders until expired otherwise. In the Taiwan case, 
orders could be canceled before they expired.
Note that contracts could by no means be bought (sold) from (to) oneself.
Figures 1 and 2 show the price time-series for each contract in the two 
experiments. 
Time is measured in minutes since midnight of January 1, 1970 UTC
(coordinated universal time).
The higher frequency of trades in the second experiment 
reflects the change in rules. Our analysis of data thus focuses on the second
experiment unless otherwise stated.
Interpretation of the price movements and  
accuracy and precision of the prediction of the time-series
on election outcomes are beyond the scope of this paper. We
briefly mention here that vote-share rankings by the
means of the price time-series
correctly mirrored the election outcomes in both experiments,
which is also true for our earlier experiment on Taiwan presidential
election in March 2004\cite{wang04}.

\section{Data Analysis}

During the tournament, information arrives stochastically and
the time intervals between successive
transactions are irregular. To generate a time series at a constant time 
interval
of 1 minute,  
we bin time into discrete values with a resolution of 1 minute. 
Prices in a time bin are then averaged. A value of zero, meaning no transactions
in that time bin, is replaced with the price in the previous time bin. 
There are thus a total of 43430 data points in such price
time series corresponding to
the duration of the tournament in minutes.
For volume time series, no such padding is performed, however.
Only 8340 nonzero data points were recorded in the volume time series.  
The ratio of 8340 to 43430 
indicates that the market was active 19\% of the time.

We call the portfolio consisting of a share of DPP, KMT, NON, PFP, and
TSU a {\it bundle}. 
Similar to a stock index which is a (weighted) sum of the stock prices
of the representative companies, we sum the five price time-series of 
the individual 
futures contracts to obtain the price time-series of a bundle. 
Figure 3 shows the time-series of the summed prices and summed trading volumes.
Hereafter, the analysis will be on such summed observables unless
otherwise stated.
\begin{figure}
\includegraphics{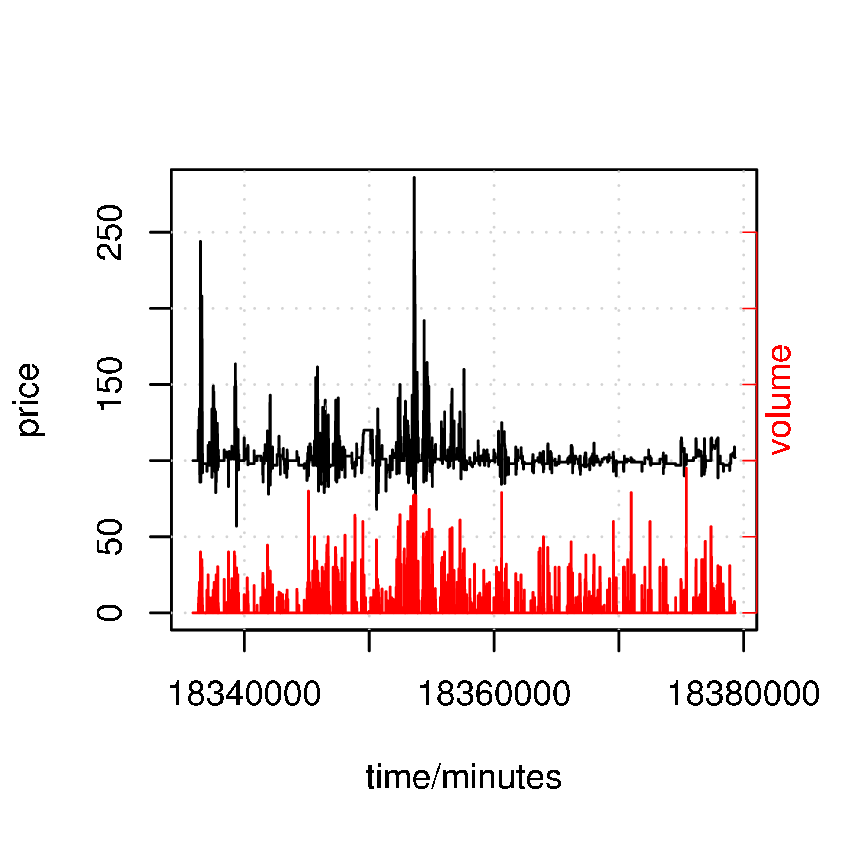}%
\caption{Time-series for the bundle price (black) and total 
volume (red) for the 2004 Taiwan parliamentary
election.}
\end{figure}


There appear many large fluctuations in Fig. 3. To study occurrence of the
fluctuations, 
we calculate the difference between the logarithmic price $\log S(t)$ 
at time $t+\tau$ and that at time $t$, 
\begin{equation}
G_\tau(t)=\log S(t+\tau) - \log S(t),
\end{equation}
and the normalized price return,
\begin{equation}
g_\tau(t) = {G_\tau(t) - \mu_\tau \over \sigma_\tau},
\end{equation}
where $\mu_\tau$ and $\sigma_\tau$ are the mean and standard deviation 
of $G_\tau(t)$. Figure 4 superposes the probability densities of the
return $g_\tau$ at
five different time lags: $\tau=$ 55, 148, 403, 1097 and 8103 minutes,
which are roughly evenly spaced in a logarithmic scale.
The scaling behavior of price returns over time lags spanning over
2 decades was well documented\cite{mandelbrot63,fama65,dacorogna93,loretan94,lux96,pagan96,bouchaud98,arneodo98,gopikrishnan99,plerou99} and reminiscent of 
the phenomena of self-organized criticality in some physical 
systems\cite{bak87,field95}. 

\begin{figure}
\includegraphics{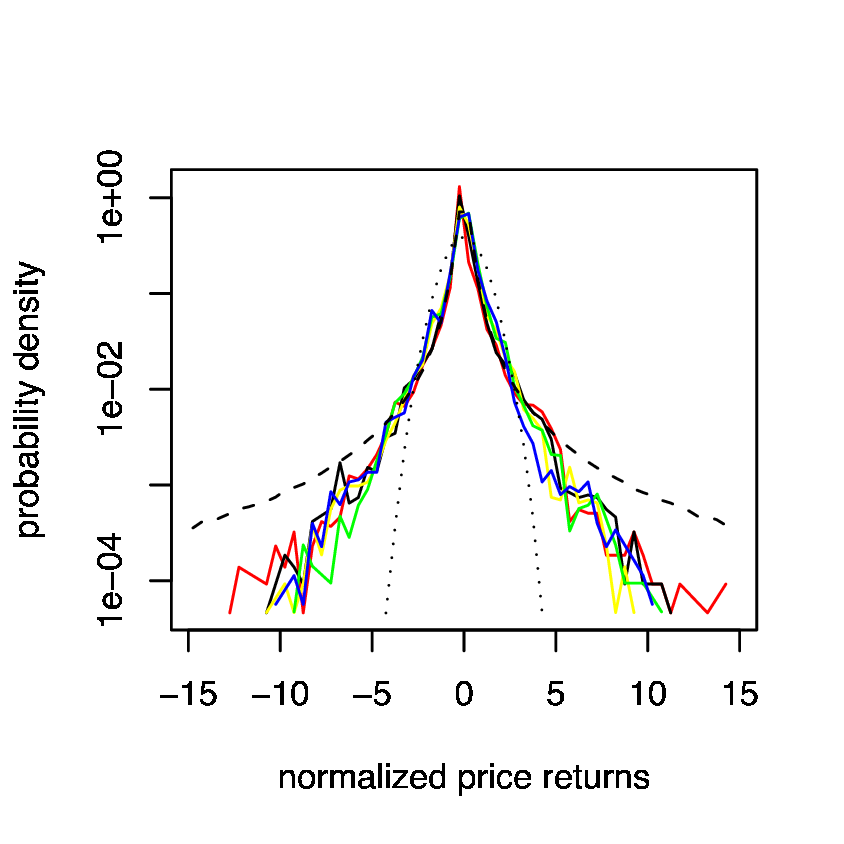}%
\caption{Probability density of normalized
price returns with time lag equal to
55 (red), 148 (black), 403 (yellow), 1097 (green)
and 8103 (blue) minutes. Dashed line is obtained from a Cauchy distribution
and dotted line a Gaussian distribution of unit variance.
}
\end{figure}

\begin{figure}
\includegraphics{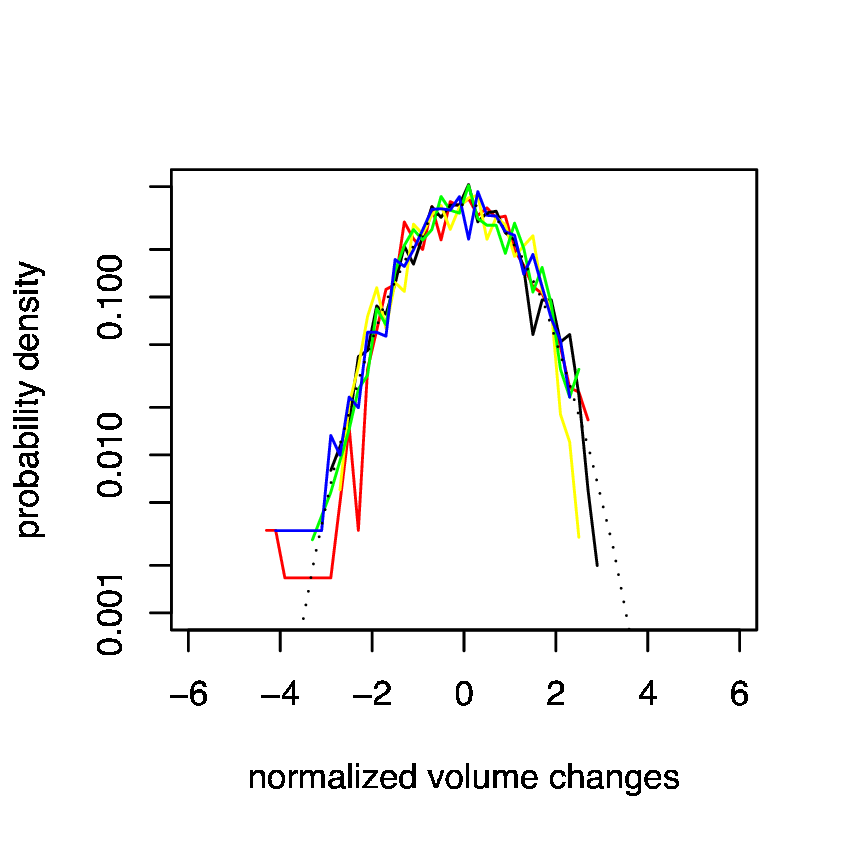}%
\caption{Probability density of normalized
volume changes with time lag equal to
55 (red), 148 (black), 403 (yellow), 1097 (green)
and 8103 (blue) minutes.
Dotted line is obtained from a standardized Gaussian distribution.
}
\end{figure}



\begin{figure}
\includegraphics{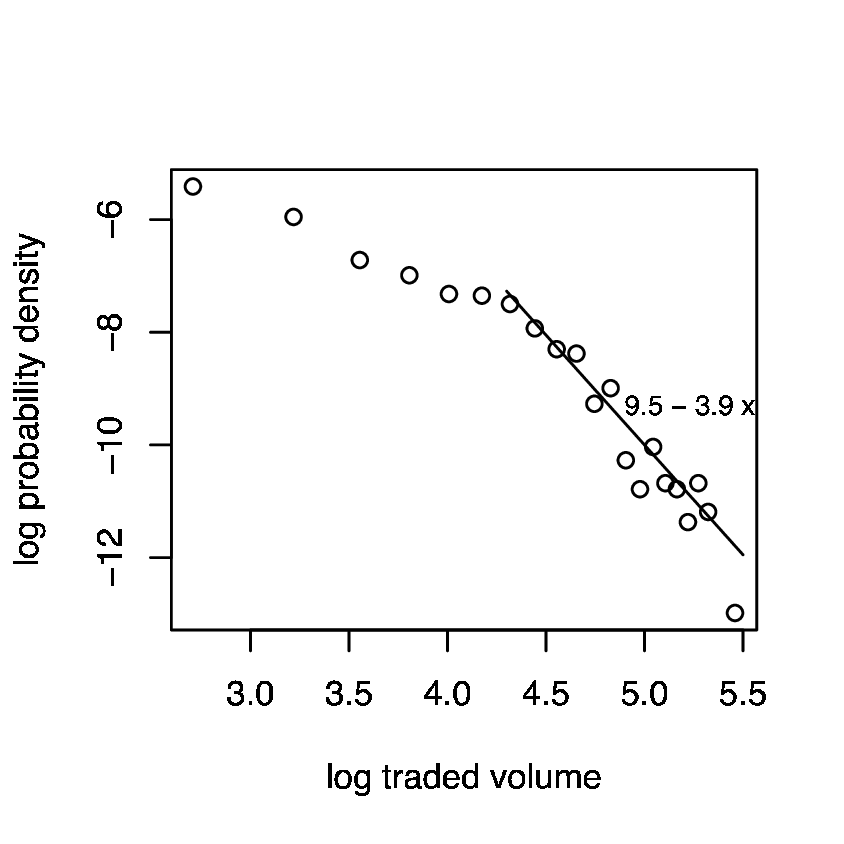}%
\caption{Probability density of volume traded (log transformed).
The straight line results from a linear regression fit to
the large volume data points, giving a slope of -3.9.
}
\end{figure}


In parallel to changes in
price, we calculated normalized volume changes and plot the probability
densities in Fig. 5, which, unlike the fat tails in Fig. 4, coincides
with a standardized normal distribution. The normality indicates that,
unlike price fluctuations,
volume fluctuations are independent\cite{bachelier00} over the range of
time lags
tested.  

Another distribution of interest is that of volumes\cite{gopikrishnan00}
which we show in Fig. 6.
A straight line fit of the distribution $p_V$ for large
trading volumes $V$ gives,
\begin{equation}
p_V \sim {1 \over V^{3.9}}\,.
\end{equation}

At the end of the tournament, we liquidated the futures contracts left
in players' accounts, the wealth of which can then be calculated.
Recall that every player was allocated an amount of 3100 
(units of fictitious money) when his account was opened. If 
the account wealth remains to be 3100 after liquidation, 
the account is deemed
inactive. To obtain the distribution of wealth, we removed inactive 
accounts, leaving 319 active ones. 
The wealth distribution $p_W$ of the active accounts 
is shown in Fig. 7, a linear fit to which suggests
a power law distribution\cite{pareto97,zipf65}, 
\begin{equation}
p_W \sim {1 \over W^{2.1}}\,.
\end{equation}

\begin{figure}
\includegraphics{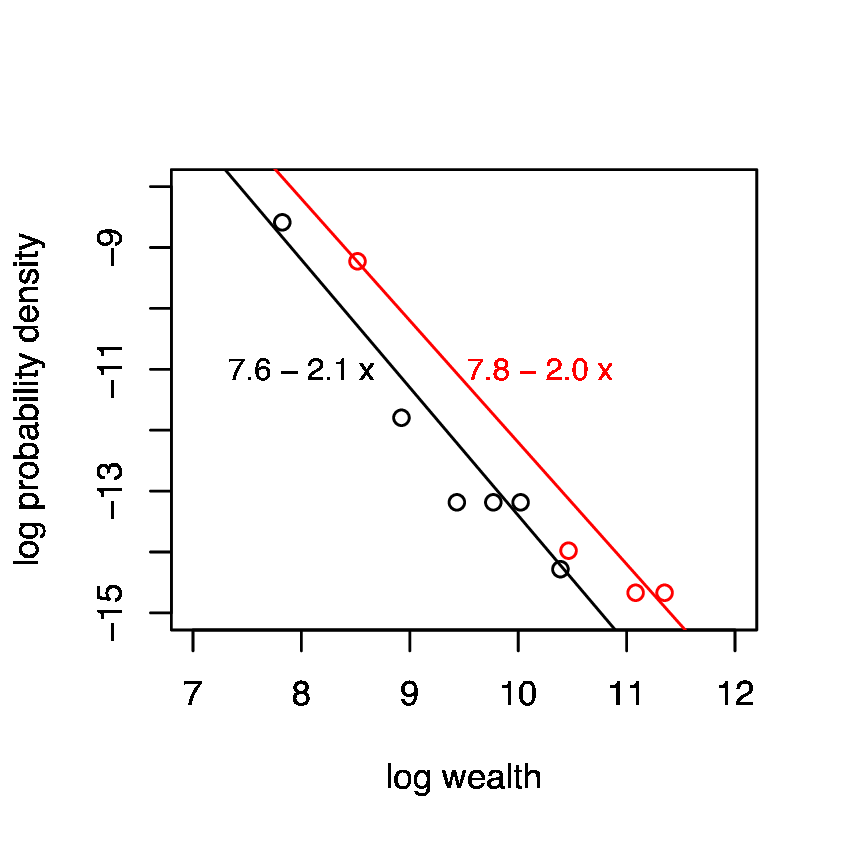}%
\caption{Probability density of wealth (log transformed).
The straight lines result from linear regression fits
to the whole range of the wealth,
giving a slope of about 2. Black and red are, respectively,
for the 2004 Taiwan parliamentary
and 2004 U.S. presidential election.
}
\end{figure}

\section{Discussion}
The fat tails in the distribution of price returns
have long been observed and suggested to be
power-law distributed\cite{mandelbrot63,fama65,dacorogna93,loretan94,lux96,pagan96,bouchaud98,arneodo98,gopikrishnan99,plerou99}. 
We performed a linear fit to the log transformed probability density of
$g_{148}$ for $g_{148} > 4$ and obtained an asymptotic density 
$p_{g_{148}}$
for the normalized returns $g_{148}$,
\begin{equation}
p_{g_{148}} \sim {1 \over {g_{148}}^{4.9}}\,.
\end{equation}
We note however that the exponent can differ if different constructions of
time series 
are used. In the above, we padded missing prices using the last
transaction price. The interpolation is valid under the assumption that
players consider the current price fair and thus do not bother to buy or sell. 
However, one can argue that the lack of 
transactions only reflects the fact that no one is online during the time bin, 
rather than an agreement on price among players. 
We therefore also analyzed the data without padding. 
In this case, 
the difference between prices at $t+\tau$ and $t$ can only be formed when
both prices exist, resulting in a drop in statistics. 
Nevertheless, Fig. 8 shows  
the scaling behavior of the (normalized) price returns thus formed $g'_{\tau}$. 
The exponent of the
positive tail of the density of $g'_{148}$ is now 3.1,
\begin{equation}
p_{g'_{148}} \sim {1 \over {g'_{148}}^{3.1}}\,.
\end{equation}

\begin{figure}
\includegraphics{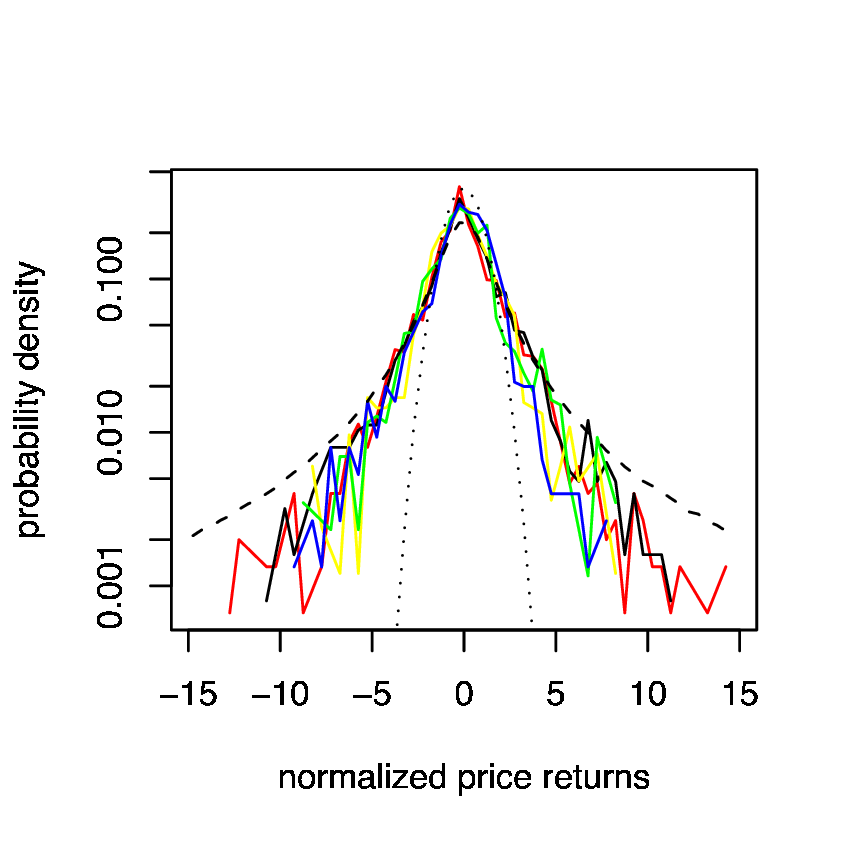}%
\caption{Probability density of normalized
returns with time lag equal to
55 (red), 148 (black), 403 (yellow), 1097 (green)
and 8103 (blue) minutes. Returns are calculated from the transaction prices. 
Dashed line is obtained from a Cauchy distribution
and dotted line a Gaussian distribution of unit variance.
}
\end{figure}

\begin{figure}
\includegraphics{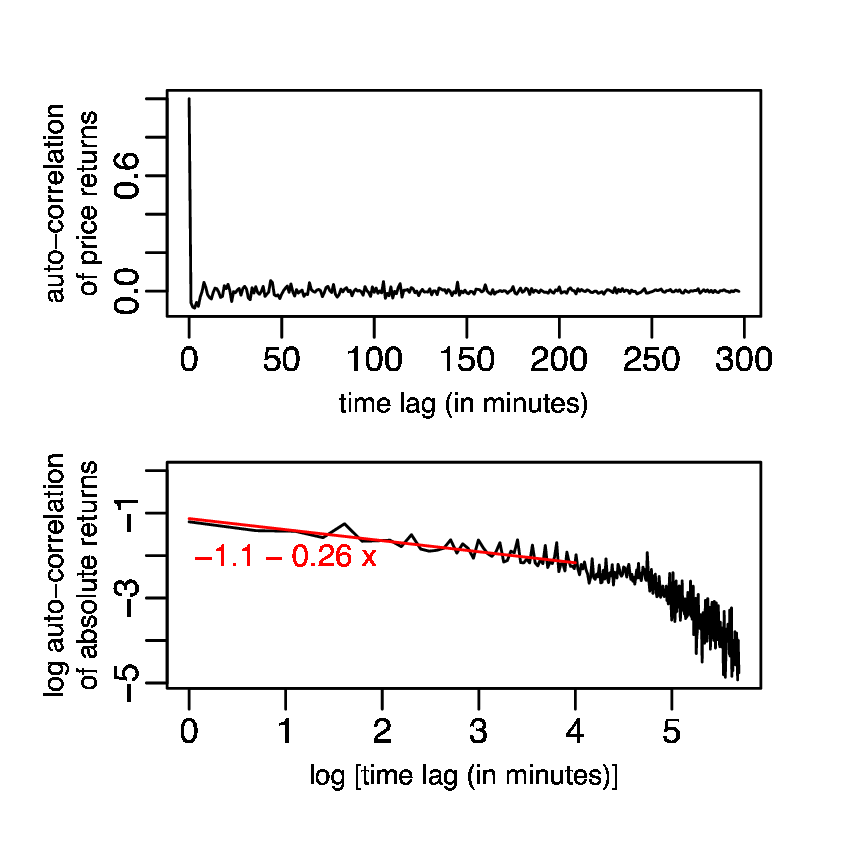}%
\caption{Autocorrelation function of $G_1(t)$ (top) and $|G_1(t)|$ (bottom).
A slope of -0.26 results from a linear regression fit.
}
\end{figure}


The exponents of the price returns are outside the stable L\'{e}vy 
regime. In Fig. 9 is plotted the autocorrelation functions of the price
returns and the absolute value of the price returns. 
Note that data are truncated at time lag equal to 298 minutes, where
the first negative autocorrelation of $|G_1(t)|$ occurs. 
It is seen that the autocorrelation of price returns drops to the noise level in
about half an hour, after which the market is considered efficient. 
Higher order correlations however persist longer, as seen in the 
slow decay of the autocorrelation of the absolute value of the price returns 
in the bottom panel of 
Fig. 9,
suggesting
that  
traders have long range memories of the magnitude of 
price changes\cite{ding83,dacorogna93,liu99}.

Our exchange server, which was open 24 hours a day 7 days a week, 
received orders
from online players who submitted their orders 
in response to random arrivals of information on campaign activities.
An order was carried out only when it intersected with a matching order before
it expired. When orders were matched, 
transaction took place. We calculated the time
intervals between successive transactions. Figure 10 shows the plot of
the numbers of transactions versus 
inter-transaction time intervals measured in minutes. It is seen
that the numbers decay asymptotically in a power-law fashion with an exponent
of 1.2. The non-exponentiality of the distribution indicates that transactions
do not take place randomly in time, even though orders are assumed 
to be submitted randomly in time.

\begin{figure}
\includegraphics{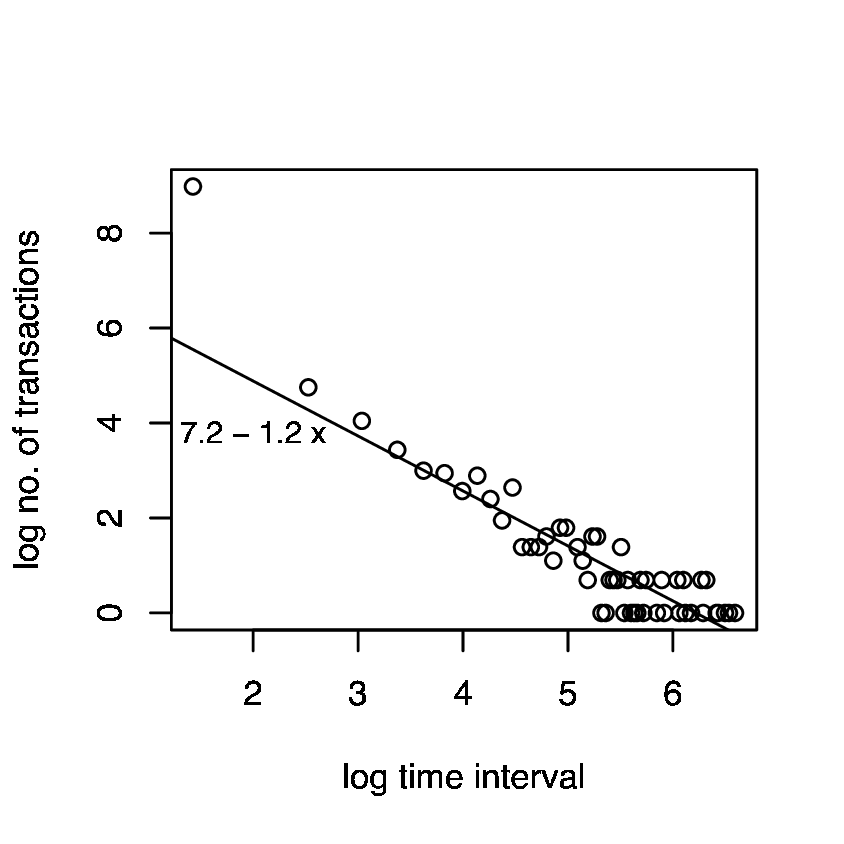}%
\caption{Distribution of inter-transaction time
intervals (log transformed).
The straight line is from a linear regression fit to
the
data points but the first one, having a slope of -1.2.
}
\end{figure}

A limit order is placed with an upper bound for buying (or
lower bound for selling) a volume of shares, 
expiring in a period of time specified by the bidder (seller).
A market order, on the other hand, buys (or sells) 
from (to) the existing orders in the orderbook, and
is executed immediately after it is received
on our server.
We can therefore say that cautious traders tend to use limit orders while 
impatient traders use more market orders.
The effect of the limit order-limit order interactions and limit order-market
order interactions 
on the price is interesting. Each futures
contract has its lowest ask and highest bid
price as a function of time. We summed the five time series to form the lowest
ask and highest bid time series of the bundle. 
In Fig. 11 is plotted the 
time series, which are seen to flank the price time series of Fig. 3.  
We calculate the arithmetic mean of the lowest ask and highest bid at any time
and obtain a time series, the   
scaling property of which is shown in Fig. 12. The spread of the
tails, compared with that in Fig. 4, does not seem to support the 
exacerbating effect of market orders. However, since the market was thin, 
players might have learned quickly to avoid placing market orders. 
More studies are needed to understand the impact of market orders.

%

In our experiment, an equal amount of money was made available to the market 
whenever a new player joined
the tournament. 
Players' money
was redistributed via trading as the tournament went on 
(a player owned on average 11 shares of each contract in the experiment).
We showed in Fig. 7 that
the distribution of wealth after the 2004 Taiwan parliamentary
election
is power-law distributed.  
In the independent experiment on the 2004 U.S. presidential election,
we also examined the wealth distribution of the active players (235 in this
case) and found a similar exponent for the power law (red line in Fig. 7). 
The Paretian property appears robust considering the lower changeover rate
of the futures contracts in the 2004 
U.S. presidential tournament than in the 2004 Taiwan
parliamentary tournament ({\it cf}. Figs. 1 and 2).
The asymptotic fat tailed distribution of price fluctuations 
also appeared in the 2004 U.S.
presidential tournament as we carried out a similar analysis
on the time series
despite their lower
statistics.
Formation of the Paretian wealth distributions could be attributed to the large 
price fluctuations, not the frequency of
trades, according to our experiment. 
To study the effect of different trading rules (social insurance policies) on 
wealth redistribution, we can, for example, charge a fee (tax)
on every transaction (income).
We can also study the dynamics by sampling the wealth distribution 
along the tournament.

\begin{figure}
\includegraphics{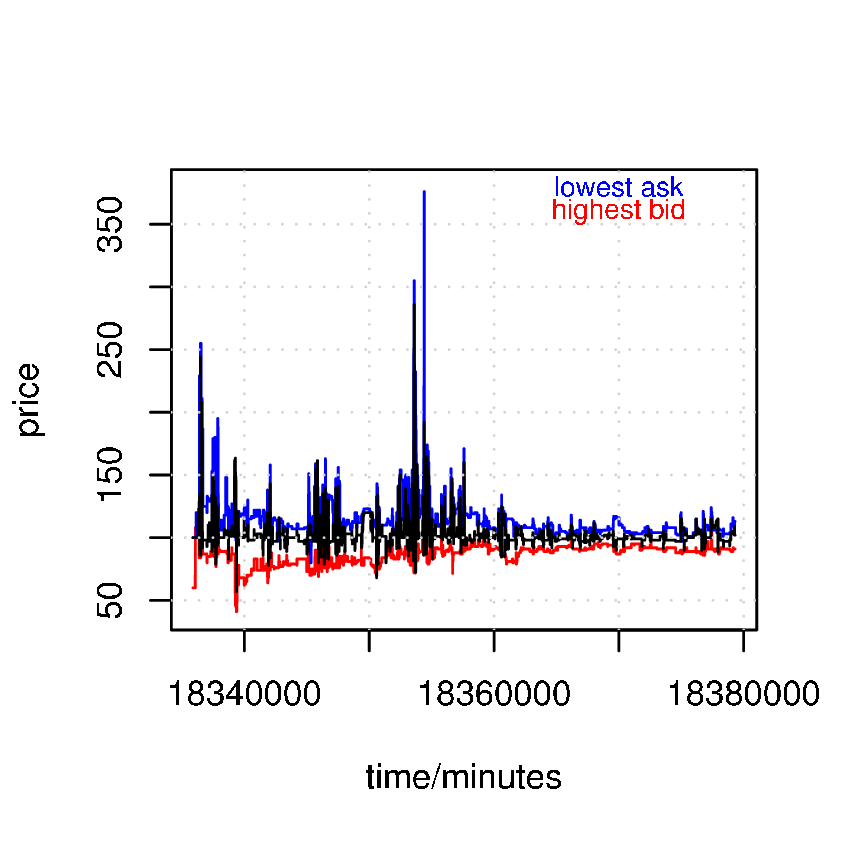}%
\caption{Lowest
ask and highest bid time-series of the 2004 Taiwan parliamentary
election.
}
\end{figure}

\begin{figure}
\includegraphics{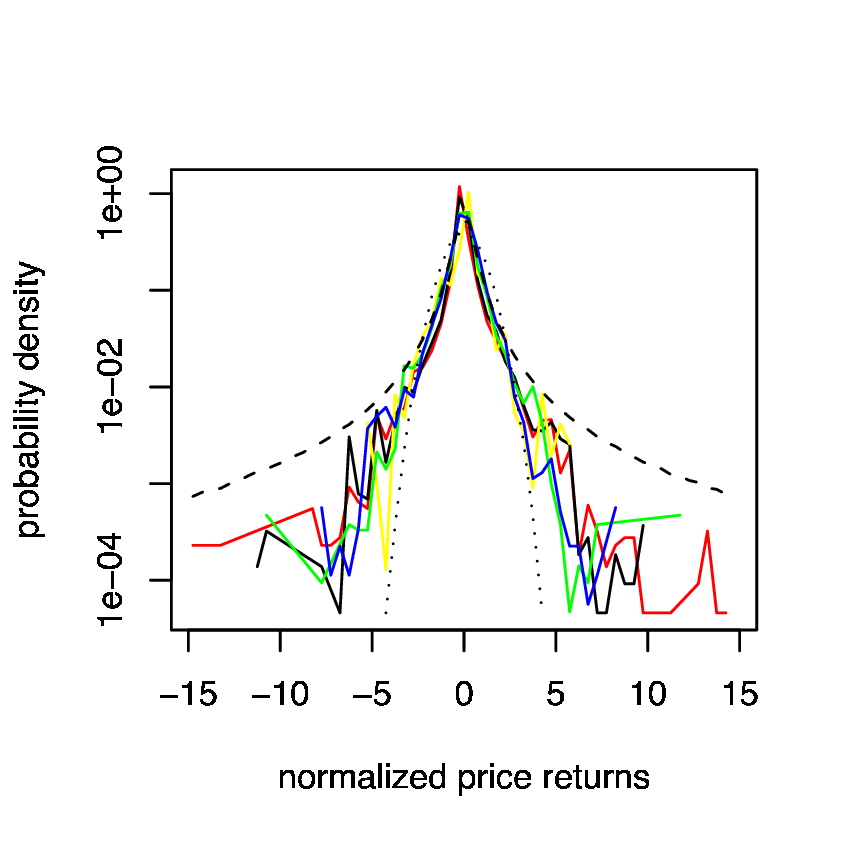}%
\caption{Probability density of normalized
price returns with time lag equal to
55 (red), 148 (black), 403 (yellow), 1097 (green)
and 8103 (blue) minutes.
Returns are calculated from
the arithmetic means of lowest asks and highest bids.
Dashed line is obtained from a Cauchy distribution
and dotted line a Gaussian distribution of unit variance.
}
\end{figure}

A simple survey of the geographical and occupational
information on the top 20 players
indicates that they do not know one another in person,
suggesting that social networks are not necessary to explain the 
power law property of the price returns and wealth.
The decay times in the price autocorrelation functions differ.
In particular, the decay time of the DPP price autocorrelation function 
is found the longest, suggesting
that there were more DPP supporters in the tournament or
that the DPP supporters were more loyal. 
Dependencies of the price changes could be caused by the 
collective actions of segments (coalitions) of participants 
of different genres, contributing to the large fluctuations.

In summary,
we have presented an approach to study the principles underlying 
complex and strongly fluctuating
socioeconomic systems. 
Futures contracts corresponding to a social event were designed. 
Futures trading experiments with well defined initial conditions  
were then set up. 
Participants were recruited and contributed to the study via the Internet.
Market observables such as transaction price, trading volume, bid
ask price, were recorded at real time.
Scaling behaviors were found in the distributions of 
price returns and trading volume similar to those found in real 
financial markets. 
Power law behaviors were also found in the distributions of inter-transaction
time intervals as well as 
participants' wealth.

\bibliography{taipex}
\end{document}